# Confinement Heteroepitaxy: Realizing Atomically Thin, Half-van der Waals Materials

Natalie Briggs[1,2,3], Brian Bersch[1,2], Yuanxi Wang[2,3], Nadire Nayir[3,4], Roland J. Koch[5,6], Ke Wang[7], Marek Kolmer[8], Wonhee Ko[8], Ana De La Fuente Duran[1], Shruti Subramanian[1,2], Chengye Dong[1,2], Jeffrey Shallenberger[7], Aaron Bostwick[5], Chris Jozwiak[5], Eli Rotenberg[5], An-Ping Li[8], Adri C. T. van Duin[1,3,4,7,9,10,11], Vincent Crespi[2,3,7,12], Joshua A. Robinson[1,2,3,7,13*]

[1]Department of Materials Science and Engineering, The Pennsylvania State University, University Park, PA 16802, United States of America
[2]Center for 2-Dimensional and Layered Materials, The Pennsylvania State University, University Park, PA 16802, United States of America
[3]2-Dimensional Crystal Consortium, The Pennsylvania State University, University Park, PA 16802, United States of America
[4]Department of Mechanical Engineering, The Pennsylvania State University, University Park, PA 16802, United States of America
[5]Advanced Light Source, Lawrence Berkeley National Laboratory, Berkeley, California 94720, United States of America
[6]The Molecular Foundry, Lawrence Berkeley National Laboratory, Berkeley, California 94720, United States of America
[7]Materials Research Institute, The Pennsylvania State University, University Park, PA 16802, United States of America
[8]Center for Nanophase Materials Sciences, Oak Ridge National Laboratory, Oak Ridge, TN 37831, United States of America
[9]Department of Chemistry, The Pennsylvania State University, University Park, PA 16802, United States of America
[10]Department of Engineering Science & Mechanics, The Pennsylvania State University, University Park, PA 16802, United States of America
[11]Department of Chemical Engineering, The Pennsylvania State University, University Park, PA 16802, United States of America
[12]Department of Physics, The Pennsylvania State University, University Park, PA 16802, United States of America
[13]Center for Atomically Thin Multifunctional Coatings, The Pennsylvania State University, University Park, PA 16802, United States of America
*Corresponding author
Email: jrobinson@psu.edu

**Three-dimensional epitaxial heterostructures are based on covalently-bonded interfaces, whereas those from 2-dimensional (2D) materials exhibit van der Waals interactions. Under the right conditions, however, material structures with mixed interfacial van der Waals and covalent bonding may be realized. Atomically thin layers formed at the epitaxial graphene (EG)/silicon carbide (SiC) interface[1] indicate that EG/SiC interfaces provide this unique environment and enable synthesis of a rich palette of 2D materials not accessible with traditional techniques. Here, we demonstrate a method termed confinement heteroepitaxy (CHet), to realize air-stable, structurally unique, crystalline 2D–Ga, In, and Sn at the EG/SiC interface. The first intercalant layer is covalently-bonded to the SiC, and is accompanied by a vertical bonding gradient that ends with van der Waals interactions. Such structures break out of plane centrosymmetry, thereby introducing atomically thin, non-centrosymmetric 2D**

allotropes of 3D materials as a foundation for tunable superconductivity,[2–5] topological states,[6] and plasmonic properties.[7,8]

The range of materials possible at the EG/SiC interface includes atomically-thin metals, which are promising materials for nanophotonics and plasmonics, due to potential for high optical sensitivity and tailorability compared to metallic thin films and nanoparticles.[7,9] Here, we explore group-III (Ga, In) and group-IV (Sn) elemental intercalation using a thermal evaporation-based technique. We show that defect engineering of graphene layers enables intercalation to the EG/SiC interface, and the underlying SiC serves as a template for intercalant crystallization. Importantly, during intercalation, the intentionally generated graphene defects are shown to heal, facilitating *ex situ* measurements without the need for post-growth passivation. Finally, the unprecedented high crystallinity of the 2D metals enables characterization of Fermi velocities in 2D metals, where 2D-Ga exhibits a Fermi velocity of approximately $2\times10^6$ m/s, exceeding that of graphene, while electron doping the graphene overlayers. The work presented here demonstrates that EG/SiC enables the creation of crystalline, 2D forms of metals which do not exist in nature; and establishes CHet as a new type of epitaxy for 2D metals.

### Discussion

Unique from traditional EG intercalation methods,[10–13] CHet (Figure 1a) uses plasma-treated EG and high-pressure (1–700 Torr) thermal evaporation to realize 2D metals. Graphene is grown from 6H-SiC and exposed to an oxygen plasma which generates graphene defects. Metallic precursors are then vaporized directly onto EG at high temperatures (>700°C), where plasma-induced graphene defects serve as entry points for atoms to intercalate to the EG/SiC interface. Graphene defects evolve throughout CHet, as seen in the x-ray photoelectron spectroscopy (XPS) C 1s core level spectra (Figure 1b). Utilization of an $O_2$/He plasma-treatment yields defective graphene layers that include C–OH and/or C–O–C, and C=O bonds, which manifest as additional peaks in the EG C 1s spectra (Figure 1b, middle),[14,15] and correlate with a 15× increase in the Raman D:G peak intensity ratio ($I_D/I_G$) (Figure 1c).[16] Upon metal intercalation, C–O bonding and the buffer layer component (Figure 1b) are no longer observed in the C 1s region (Figure 1b, top), and a metallic Ga 3d peak is detected (Figure S1). These observations indicate that plasma-induced C–O bonding does not persist through the final step of CHet and that the buffer layer is released from underlying SiC by an interfacial Ga layer (Figure S1, S2).[10–12] Interestingly, regardless of the metal intercalated, CHet yields a 5× decrease in $I_D$, while $I_G$ and $I_{2D}$ increase by 3–4× and 2–5×, respectively (Figure 1c, top 3 spectra). The decreased $I_D$ and lack of D′ and D+G Raman modes in conjunction with the loss of C–O bonding signature and air-stability of resulting 2D metals[17] suggest the graphene is healed, possibly due to metal-catalyzed graphene regrowth during intercalation.[18–20] Enhancement of $I_G$ and $I_{2D}$ following intercalation could be attributed to metal/graphene charge transfer or plasmonic resonance.[21–23]

Oxygen-passivated graphene defects promote metal adhesion to and intercalation through graphene. We utilize density functional theory (DFT) to reveal how carbon monovacancies and their complexes (up to octa-vacancies), either unpassivated or passivated by =O, –O–, or –OH groups, affect metal adsorption to and intercalation through graphene (See Figure S3 for details). Passivation chemistries are guided by C 1s spectra (Figure 1b), which alone cannot distinguish between C–O–C and C–OH. The adsorption energy of a Ga atom to defective graphene is found through $E_{ads} = E_{Ga+graphene} − (E_{graphene} + E_{Ga})$, where $E_{graphene+Ga}$ is the total energy of the defective graphene with a Ga atom adsorbed, and $E_{graphene}$ and $E_{Ga}$ are the energies of the two components isolated. In general, Ga bonding is strengthened with increasing vacancy size, where large, un-passivated defects bond covalently with Ga (Figure 1d, S4). Passivated vacancies, however, are more suitable to Ga intercalation. Oxygen passivation (=O and –O–) of carbon vacancy edge atoms weakens the binding strength of Ga atoms to the defect by >50% (Figure 1e,f, S4(w-II, y-I)), to binding energies low enough that Ga can be more easily released into the EG/SiC interface. –OH passivated defects (Figure 1g) are less likely to participate in the attraction and intercalation of Ga through a graphene defect due to low resulting Ga binding energies, (-2.26 eV in Figure 1g, and as low as -1.5 eV for –OH passivated tri- and penta-vacancies compared to -1.75 eV for Ga binding to pristine graphene) (Figure S4(y-II)).

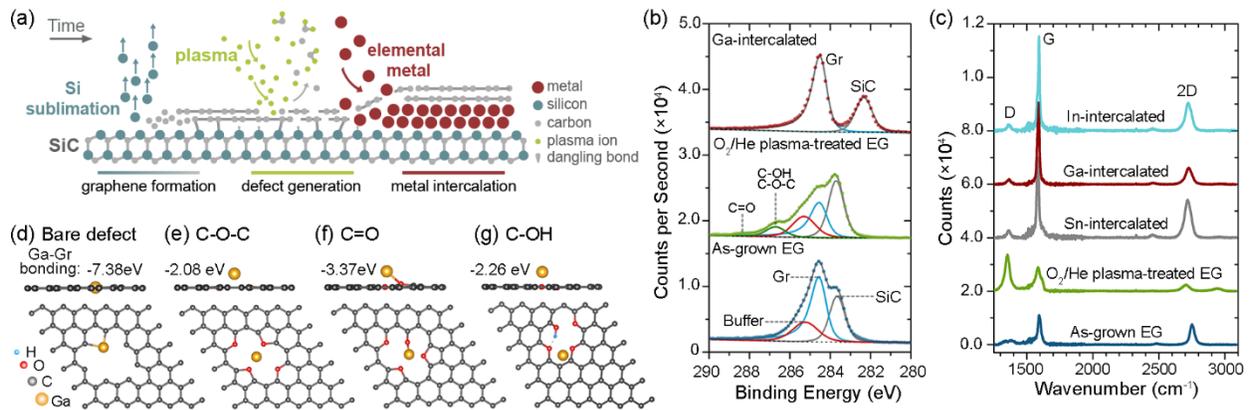

**Figure 1: Confinement Heteroepitaxy.** (a) Schematic of CHet showing EG growth, O$_2$/He plasma treatment, and intercalation steps (b) XPS showing C 1s for (bottom) as-grown EG, (middle) O$_2$/He plasma-treated EG, and (top) Ga-intercalated EG demonstrating the creation and antihalation of C-O bonds during CHet; This is confirmed by (c) Raman spectroscopy of as-grown EG, O$_2$/He plasma-treated EG, and metal-intercalated EG, where the defect peak (D peak) is dramatically reduced as a result of the intercalation process. (d)-(g) DFT modelling of Ga atoms on optimized graphene sheets with the bare, C-O-C, C=O, and C-OH, passivated defects suggests that oxygen termination (e,f) provides favorable energies for metal attraction and intercalation through the graphene sheet. The Ga binding energy to each defect is shown in each model.

The intercalated, interfacial 2D metals are 1–3 atomic layers thick, and highly registered to the SiC substrate (Figure 2a,c, e). The dominant thickness is readily described by first-principles layer phase stability calculations (Figure 2b,d,f, S7), which predict the equilibrium layer thickness (for various interlayer stacking registries) as a function of accessible metal chemical potential. These calculations yield a stability range of 1–3 layers for Ga, 2–3 layers for In, and 1 layer for Sn, in good agreement with scanning transmission electron microscopy (STEM) where the dominant experimentally observed layer numbers are 2–3 for Ga, 2 for In, and 1–2 for Sn. We note that select sample regions show different layer numbers beyond those predicted by DFT (Figure S8). Additionally, in the case of Sn, the blurred image suggests a loss of registry and metastability of the second layer compared to the first.

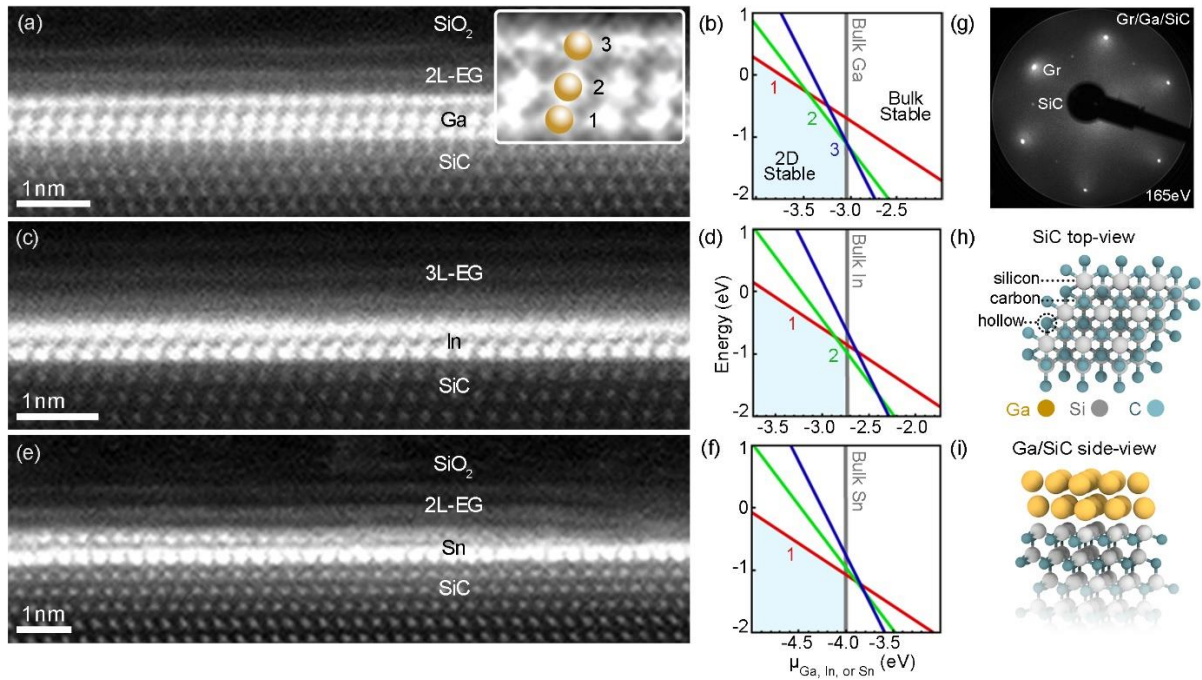

**Figure 2: Atomic Structure of CHet-grown 2D metals.** (a) Cross-sectional STEM showing 3 layers of Ga between EG and SiC, and (b) corresponding energy minimization calculations showing preferred intercalant layer numbers. Energy dispersive x-ray spectroscopy (Figure S5) confirms that intercalant layers match the respective precursor elements, and are not oxidized. The grey, vertical lines in b,d,f indicate the metal chemical potential of the bulk metal, and red, green, and blue lines indicate energy as a function of chemical potential for 1, 2, and 3 layers, respectively. The lowest line at a given potential indicates the ground state layer-number. (c-f) show cross-sectional STEM and layer number calculations for In and Sn (g) Acquired LEED pattern for Gr/Ga/SiC showing Gr and SiC spots. Patterns for EG/SiC, Gr/In/SiC, and Gr/Sn/SiC are shown in S6 (h) Top-down view of hexagonal SiC with different sites labeled. (i) Modeled 2L Ga on SiC, showing the orientation of intercalated Ga layers with the SiC substrate.

The epitaxial relationship of the 2D metal to the SiC substrate is supported by low-energy electron diffraction (LEED) patterns, which exhibit graphene and SiC spots, but lack spots corresponding to a structurally unique intercalant layer (Figure 2g, S6). This observation, in conjunction with an observed lateral atomic Ga spacing of 2.72Å (matching that of the underlying SiC Figure 2a, S11), and computational optimization of the ground state of the Ga structure establishes that the Ga lattice structure is matched to SiC. First-principles DFT is performed to investigate one- to three-layer Ga atoms initialized at sites projecting onto the silicon site, carbon site, and hollow site of SiC (Figure 2h). Adding a top bilayer graphene only affects band fillings (discussed below) without changing the relative stability of the Ga structures, therefore it is not considered further in the stability calculations (see Table S1 for cases including graphene). Following full relaxation, the ground state for one layer of Ga on SiC contains Ga located above the silicon site ($Ga_{Si}$) (Figure 2i). The second layer Ga sits at the C site ($Ga_C$) (Figure 2i), and the third layer Ga sits at the hollow site ($Ga_{hollow}$). Thus the ground state configurations for one-, two-, and three-layer Ga on SiC are $Ga_{Si}$, $Ga_{Si}/Ga_C$, and $Ga_{Si}/Ga_C/Ga_{hollow}$. This "ABC" stacking resembles a face-centered-cubic (FCC) lattice cleaved along (111), which matches the hexagonal arrangement of SiC (0001), and may be related to high-pressure, metastable and distorted FCC phases of Ga-III.[24] Comparing with other metastable structures reveals that the Ga registry weakens for increased Ga thickness: the $Ga_{Si}$ stacking site for single layer Ga is 0.14 eV more stable than $Ga_C$ and $Ga_{hollow}$, while ground states for bilayers and trilayers are only preferable against the their respective next lowest-energy competing phases ($Ga_{Si}/Ga_{Hollow}$ for bilayer, $Ga_{Si}/Ga_C/Ga_{Si}$ and $Ga_{Si}/Ga_C/Ga_C$ for trilayer) within 0.05 eV. Cross-sectional STEM supports this change in registry with increasing thickness, where interlayer spacing between $1^{st}$ and $2^{nd}$ Ga layers is smaller than between $2^{nd}$ and $3^{rd}$ (Figure 2a, inset) by nearly 10% (2.19 Å versus 2.36 Å). Some lateral translations in the bottom layer Ga (relative to top layer Si) and the third layer Ga (relative to the middle layer Ga) are also evident in STEM, which can be due to the weaker Ga registry for the second and third layer trapping Ga layers at metastable structures, or to kinetic factors, such as Ga layers stitching with nearby SiC step edges that force subsequent Ga atoms to take the $Ga_{Si}$ site. Alternative to the above thermodynamic analysis, the dominant stacking order of Ga can be identified as the structure whose DFT band structure best matches the ARPES-measured one (Figure 3b, S12, S13), since the band structure is sensitive to Ga stacking order. The dominant phase found from this method is $Ga_{Si}/Ga_C$ for bilayer Ga (Figure 2i) and $Ga_{Si}/Ga_C/Ga_C$ for trilayer, consistent with previous thermodynamic analysis of $Ga_{Si}/Ga_C$ being the bilayer ground state and $Ga_{Si}/Ga_C/Ga_C$ being a low-energy trilayer structure nearly degenerate with the ground state (and possibly being favored by kinetic factors). Matching DFT (PBE) results to ARPES bands for the bilayer case requires a Fermi level upshift of 0.6 eV (purple dashed line in Fig. 3b), however this artificial upshift is not needed when using hybrid functionals (Figure S13), which generally yield more accurate band alignments. In experiment, electron doping may arise from co-existing trilayer regions with smaller work functions.

The small atomic radii (large Brillouin zone (BZ)) and high valence electron count of the early-period $p$-block metals make them ideal potential candidates for free-electron-like metals with the largest Fermi velocities. This is found to be the case for 2D Ga by further inspection of the calculated and ARPES-measured band structure. Figure 3a shows the measured Gr/Ga ARPES band structure, where the Ga $s$-band (see below) Fermi velocity of $2\times10^6$ m/s is similar to that calculated for bulk Al and Ga in a free electron model (both $\sim 2\times10^6$ m/s),[25] and that measured for 2D indium on Si(111).[26] By comparison, the nearly linear bilayer graphene bands, shown with the highest intensity, have Fermi velocities of $1.2\times10^6$ m/s.[27,28] The location of the graphene Dirac point 0.2-0.3 eV below the Fermi level indicates the intercalation of Ga leads to $8-10\times10^{12}$ cm$^{-2}$ electron doping without introducing other hybridizations. The ARPES-measured Fermi surface (Figure 3c, d) shows circular contours from Ga, indicating nearly-free electron behavior, along with Dirac points from bilayer graphene. This is supported by the DFT-calculated Fermi surface of bilayer Ga/SiC system without graphene (Figure 3e), where the Fermi level again upshifted by 0.6 eV to be consistent with the ARPES measured band alignment.

The calculated band structure (Figure 3b) shows projection of the total wavefunction onto the plane-wave components of the graphene (black) and Ga/SiC (blue) primitive cell, where effective band structures are unfolded from the BZ of the supercell.[29] The calculated bilayer Ga band structure agrees with the measured ARPES data along the $\Gamma M_{Ga}$ ($\Gamma K_g$) and $\Gamma K_{Ga}$ ($\Gamma M_g$) directions (Figure 3b inset). The most prominent features contributed by Ga are three avoided band crossing points, one along $\Gamma M_{Ga}$ and two along $\Gamma K_{Ga}$. To reveal the orbital origin of the band crossing along $\Gamma M_{Ga}$, we compare the projected band structure of bilayer Ga/SiC (without graphene) to a hypothetical freestanding bilayer Ga where Ga atoms are frozen at their positions in the hybrid system (Figure S14). The latter shows three nearly-free-electron-like bands of $s$-bonding, $s$-antibonding, and $p$ orbital character. The band crossing is thus hybridization between a parabolic $s$ orbital originating from ~9 eV below the Fermi level and the p orbital near the Fermi level. Similarly, the band crossings along $\Gamma K_{Ga}$ are also between $s$ and $p$.

The measured Fermi surface also shows that the graphene BZ is oriented 30° rotated from the underlying Ga/SiC BZ, further supporting epitaxial relation between Ga and SiC. In addition to Ga/SiC, ARPES measurements are performed for In/SiC. The resulting measurements resemble those of Ga/SiC in Figure 3, where the sample exhibits graphene bands near $K_g$, in addition to avoided band crossing points of In along $\Gamma M_{Ga}$ and $\Gamma K_{Ga}$. Additionally, the graphene BZ zone is 30° rotated from that of the In/SiC BZ (Figure S15).

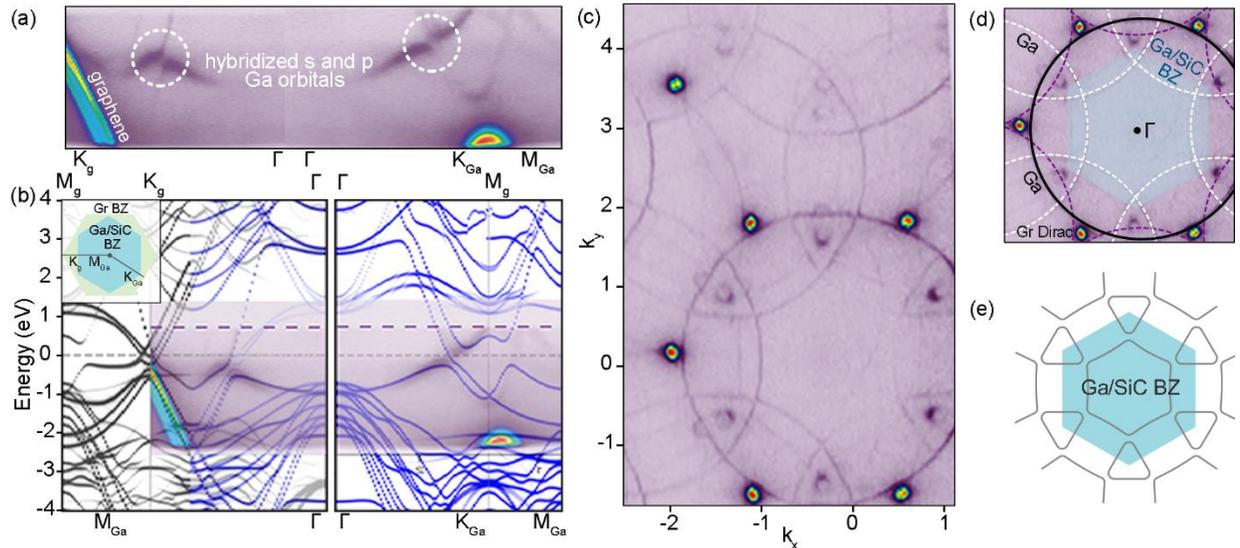

**Figure 3: Electronic Structure of CHet-grown 2D-Ga.** (a) Acquired ARPES for Gr/Ga/SiC showing graphene and Ga bands near $M_{Ga}$ and $K_{Ga}$ and (b) Effective unfolded band structures of 2×2 graphene + √3×√3 R30° bilayer Ga/SiC along the $\Gamma M_{Ga}$ ($\Gamma K_g$) and the $\Gamma K_{Ga}$ ($\Gamma M_g$) directions, as indicated by the Brillouin zone paths in the inset. ARPES measurements along the same paths are superimposed. (c, d) ARPES measured Fermi surface showing nearly-free-electron-like circular contours. Black, white, and purple circles in (d) correspond to nearly free electron like circular contours from Ga, and are drawn to aid in comparison of experimental data with (e) DFT-calculated Fermi surface of bilayer Ga/SiC with the Fermi level shifted to match the measured band filling.

## Conclusion

Confinement Heteroepitaxy stabilizes 2D forms of 3D materials, resulting in hybrid bonding that creates strong symmetry breaking across the interface to enable unique phenomena not found in bulk counterparts. Importantly, overlying graphene layers utilized in CHet not only help confine the 2D metal, but also serve as a hermetic seal to preclude oxidation of ultrathin non-noble metals. The ability to perform extensive *ex situ* characterization of these materials reveals the robust nature of the Gr/intercalant/SiC structure. As a result, this work opens the door to the study of new heterostructures based on 2D-metal layers, as well as investigations of intrinsic 2D-metal layer properties, including plasmonic behavior and superconductivity.

## Acknowledgements

Funding for this work was provided by the 2D Crystal Consortium National Science Foundation (NSF) Materials Innovation Platform under cooperative agreement DMR-1539916, the Semiconductor Research Corporation Intel/Global Research Collaboration Fellowship, task 2741.001, Northrop Grumman Mission Systems' University Research Program, NSF CAREER Award: 1453924, and the Chinese Scholarship Council. A portion of this research was conducted at the Center for Nanophase Materials Sciences, which is a DOE Office of Science User Facility, and at the Pennsylvania State University Materials Research Institute's Material Characterization Laboratory. This research used resources of the Advanced Light Source, which is a DOE Office of Science User Facility under contract no. DE-AC02-05CH11231. A portion of this work was also supported by the National Science Foundation DMR-1808900.

## Methods

*Epitaxial graphene synthesis*
Epitaxial graphene synthesis was performed according to methods described in related manuscripts (SI ref. 6, 7)

*Epitaxial graphene plasma treatment*
Epitaxial graphene layers were plasma treated using a Tepla M4L plasma etch tool, using 150 sccm $O_2$ and 50 sccm under a pressure of 500 mTorr and power of 50 W. Additional plasma chemistries ($CF_4$ and $H_2/N_2$) have been explored but not investigated in depth.

*2D-Metal Intercalation*
Metal intercalation was performed using an STF-1200 horizontal tube furnace fitted with a 1" O.D. quartz tube. A custom-made alumina crucible from Robocasting Enterprises was used to hold 1x1cm EG/SiC substrates, which were placed with graphene layers on the Si face of SiC facing downward, toward the inside of the crucible. 30-60 mg of metallic Ga (Sigma Aldrich, 99.999%), In powder (Alfa Aesar, -325 mesh, 99.99%), or Sn granules (Alfa Aesar, 99.5%) were placed in the crucible, directly beneath the EG/SiC substrate. The crucible with EG/SiC and the respective metal precursor was then loaded into the tube furnace and evacuated to ~5 mTorr. The tube was evacuated for a sufficient amount of time (~30 minutes) to ensure the pressure rate-of-rise over 5 minutes did not exceed 5 mTorr/minute. The tube was then pressurized to 300 Torr with Ar. At this time, the furnace was heated to 600-800°C under a ramp rate of 20°/minute. The furnace was held at maximum temperature for 30 minutes, then cooled with a fan to approximately 30°C. An Ar flow of 50 sccm was maintained throughout the heating process.

*X-ray photoelectron spectroscopy*
X-ray photoelectron spectroscopy measurements were carried out with a Physical Electronics Versa Probe II equipped with a monochromatic Al $K_\alpha$ X-ray source (hv=1486.7 eV) and a concentric hemispherical analyzer. High resolution spectra were obtained over an analysis area of 200 μm at a pass energy of 29.35 eV for C 1s, Si 2p, Ga 3d, and Ga 2p regions. O 1s regions were collected with a pass energy of 46.95 eV. The acquired spectra were fitted Lorentzian lineshapes, and the asymmetric graphene peak fit was derived from exfoliated highly oriented pyrolytic graphite and H-intercalated epitaxial graphene reference samples.

Spectra were charge referenced to this graphene peak in C 1s corresponding to 284.5 eV. A U 2 Tougaard background was used to fit XPS spectra.

*Raman spectroscopy*
Raman spectra were acquired with a Horiba LabRam Raman system using a wavelength of 488nm and a power of 4.6 mW. Spectra are acquired with an integration time of 30s, using a 600 grooves/mm grating.

*Cross-sectional transmission electron microscopy*
Cross-sectional samples for STEM imaging were prepared by in-situ lift-out via milling in a FEI Helios NanoLab DualBeam 660 focused ion beam (FIB). Prior to FIB, 60/5/10 nm of $SiO_2$/Ti/Au was deposited via electron-beam evaporation in a Kurt J. Lesker Lab18 evaporator, to improve contrast during STEM imaging at low magnifications. Cross-sections were prepared using a Ga+ ion beam at 30 kV then stepped down to 1 kV to avoid ion beam damage to the sample surface.

High resolution scanning transmission electron microscopy (STEM) of sample cross sections was performed in a FEI dual aberration corrected Titan3 G2 60-300 S/TEM at 200kV using a high angle annular dark field (HAADF) detector. The HAADF detector (Fischione) has a collection angle of 51-300 mrad for Z-contrast imaging. A beam current of 70pA, beam convergence of 30 mrad (C2 aperture of 70 μm), and camera length of 115 mm are used for STEM image acquisition. The STEM EDS maps are collected by using the superX EDS system, which has 4 EDS detectors surrounding the sample.

*Low-energy electron diffraction*
Low-energy electron diffraction measurements of Gr/Ga/SiC, Gr/In/SiC, and Gr/Sn/SiC samples were performed using LEED Spectrometer BDL800IR-MCP manufactured by OCI Vacuum Microengineering. Samples were first degassed at 200°C for 30 minutes under UHV to desorb surface moisture and contaminants. LEED patterns were then acquired at room temperature using constant primary beam currents of 10 nA and beam energies of 50 eV – 250 eV, in 1 eV steps.

*Angle-resolved photoemission spectroscopy*
Angle-resolved photoemission spectroscopy measurements were performed at the Microscopic and electronic structure observatory (MAESTRO) beamline at the Advanced Light Source at Lawrence Berkeley National Lab. The sample was annealed at 550 K for 30 minutes in the end-station before measurements to remove adsorbates from the transfer of the sample through air. Measurements of Gr/Ga/SiC and Gr/In/SiC structures were performed using a photon energy of 140 eV and 110 eV, respectively. Photoemission spectra were collected by moving the sample around one angle while using the angle resolved mode of a Scienta R4000 electron analyzer for the collection of the other angular axis.

*Theory*
i.  Graphene Defect Generation and Passivation, Ga Adsorption
    All density functional theory calculations investigating the role of plasma treatment on EG defects and Ga intercalation were performed in Quantum Espresso (SI ref. 8), using projected augmented wave pseudopotentials (SI ref. 9, 10) and the Perdew-Burke-Ernzerhof parametrization of the generalized gradient approximation exchange-correlation functional (GGA-PBE, SI ref. 11, 12). A 5×5×1 Γ-centered k-point mesh was applied for Brillouin zone integration. Planewave expansions were truncated at an energy cut-off of 408 eV for wavefunctions and at 4080 Ry for charge densities. The Marzari-Vanderbilt cold smearing scheme (SI ref. 13) was applied with a broadening of 0.1 eV. Structural relaxations used the Broyden–Fletcher–Goldfarb–Shanno algorithm with a force threshold of 0.025 eV/Å. A vacuum layer of 20 Å was inserted in the direction normal to the graphene sheets to minimize the spurious interactions across the periodic boundary. The models in the figures were visualized using OVITO (SI ref. 14 ) and VESTA (SI ref. 15)  software.

ii.  2D Ga phase stability and electronic structure calculations

All density functional theory calculations on phase stabilities and electronic structure were performed using the GGA-PBE exchange-correlation functional (SI ref. 11) and the projector augmented wave (PAW) pseudopotentials (SI ref. 9, 10). Plane-wave expansions were truncated at an energy cutoff of 500 eV. All structural relaxations were performed using dipole corrections to the total energy (SI ref. 16) and to the electrostatic potential (SI ref. 17) in the out-of-plane direction, until the remaining forces are within 0.01 eV/Å. All Ga/SiC calculations were performed using 7 repeating units of SiC along the $z$ direction as substrate, capped by Ga from above and by hydrogen from below. Graphene/Ga/SiC calculations were performed using 5 repeating units along the $z$ for the a 2×2 graphene + √3×√3 R30º Ga/SiC supercell, and 3 repeating units for the a 5×5 graphene + 4×4 R0º Ga/SiC supercell to alleviate the computational demand of accommodating more atoms in the large supercells. Band unfolding were performed using the GPAW package (SI ref 18); all other calculations were performed by the Vienna Ab Initio Package (VASP) (SI ref 19). Fermi surfaces of Ga/SiC are calculated on a 40×40×1 grid and interpolated onto a 200×200×1 grid for plotting. Band structures at the hybrid functional level were calculated using the range-separated form of Heyd, Scuseria, and Ernzerhof (SI ref. 20) (HSE06, i.e. with a range-separation parameter of 0.2 Å$^{-1}$) and using structures relaxed at the PBE level. Self-consistency HSE06 calculations were performed on a 12×12×1 k-point grid.

## References

1. Al Balushi, Z. Y. *et al.* Two-dimensional gallium nitride realized via graphene encapsulation. *Nat. Mater.* **15**, 1166–1171 (2016).

2. Gregory, W. D. Superconducting Transition Width in Pure Gallium Single Crystals. *Phys. Rev.* **165**, 556–561 (1968).

3. Xing, Y. *et al.* Quantum Griffiths singularity of superconductor-metal transition in Ga thin films. *Science (80-. ).* **350**, 542–545 (2015).

4. Zhang, T. *et al.* Superconductivity in one-atomic-layer metal films grown on Si(111). *Nat. Phys.* **6**, 104–108 (2010).

5. Liao, M. *et al.* Superconductivity in few-layer stanene. *Nat. Phys.* **14**, (2018).

6. Deng, J. *et al.* Epitaxial growth of ultraflat stanene with topological band inversion. *Nat. Mater.* **17**, 1081–1086 (2018).

7. Boltasseva, A. & Shalaev, V. M. Transdimensional Photonics. *ACS Photonics* **6**, 1–3 (2019).

8. Maniyara, R. A. *et al.* Tunable plasmons in ultrathin metal films. *Nat. Photonics* **13**, 328–333 (2019).

9. Briggs, N. *et al.* Epitaxial Graphene Intercalation: A Route to Graphene Modulation and Unique 2D Materials. *arXiv:1905.09261* (2019).

10. Riedl, C., Coletti, C. & Starke, U. Structural and electronic properties of epitaxial Graphene on SiC(0001): A review of growth, characterization, transfer doping and hydrogen intercalation. *J. Phys. D. Appl. Phys.* **43**, (2010).

11. Emtsev, K. V, Zakharov, A. A., Coletti, C., Forti, S. & Starke, U. Ambipolar doping in quasifree epitaxial graphene on SiC (0001) controlled by Ge intercalation. *Phys. Rev. B* **84**, 1–6 (2011).

12. Gierz, I. *et al.* Electronic decoupling of an epitaxial graphene monolayer by gold intercalation. *Phys. Rev. B* **81**, 1–6 (2010).

13. Virojanadara, C., Watcharinyanon, S., Zakharov, A. A. & Johansson, L. I. Epitaxial graphene on 6H-SiC and Li intercalation. *Phys. Rev* **82**, 1–6 (2010).

14. Moulder, J. F. & Chastain, J. *Handbook of x-ray photoelectron spectroscopy : a reference book of standard spectra for identification and interpretation of XPS data*. (Physical Electronics Division, Perkin-Elmer Corp, 1992).

15. Beamson, G. (Graham) & Briggs, D. (David). *High resolution XPS of organic polymers : the Scienta ESCA300 database*. (Wiley, 1992).

16. Eckmann, A. *et al.* Probing the Nature of Defects in Graphene by Raman Spectroscopy. *Nano Lett.* **12**, 3925–3930 (2012).

17. Bersch, B. *et al.* An Air-Stable and Atomically Thin Graphene/Gallium Superconducting Heterostructure. *https://arxiv.org/abs/1905.09938* (2019).

18. Vishwakarma, R. *et al.* Transfer free graphene growth on SiO2 substrate at 250 °C. *Sci. Rep.* **7**, 43756 (2017).

19. Araby, M. I. *et al.* Graphene formation at 150°C using indium as catalyst. *RSC Adv.* **7**, 47353–47356 (2017).

20. Fujita, J. *et al.* Near room temperature chemical vapor deposition of graphene with diluted methane and molten gallium catalyst. *Sci. Rep.* **7**, 12371 (2017).

21. Yi, C. *et al.* Evidence of Plasmonic Coupling in Gallium Nanoparticles / Graphene / SiC. *Small* **8**, 2721–2730 (2012).

22. Losurdo, M. *et al.* Demonstrating the Capability of the High-Performance Plasmonic Gallium À Graphene Couple. 3031–3041 (2014). doi:10.1021/nn500472r

23. Khorasaninejad, M. *et al.* Highly Enhanced Raman Scattering of Graphene using Plasmonic Nano-Structure. *Sci. Rep.* **3**, 1–7 (2013).

24. Voloshina, E., Rosciszewski, K. & Paulus, B. First-principles study of the connection between structure and electronic properties of gallium. *Phys. Rev. B* **79**, 045113 (2009).

25. Ashcroft, N. W. & Mermin, N. D. *Solid state physics*. (Holt, Rinehart and Winston, 1976).

26. Yoshizawa, S., Kim, H., Hasegawa, Y. & Uchihashi, T. Disorder-induced suppression of superconductivity in the Si(111)-(7 × 3) -In surface: Scanning tunneling microscopy study. *Phys. Rev. B - Condens. Matter Mater. Phys.* **92**, 041410 (2015).

27. Yang, L., Deslippe, J., Park, C.-H., Cohen, M. L. & Louie, S. G. Excitonic Effects on the Optical Response of Graphene and Bilayer Graphene. *Phys. Rev. Lett.* **103**, 186802 (2009).

28. Zhang, Y., Tan, Y.-W., Stormer, H. L. & Kim, P. Experimental observation of the quantum Hall effect and Berry's phase in graphene. *Nature* **438**, 201–204 (2005).

29. Popescu, V. & Zunger, A. Extracting $E$ versus $\vec{k}$ effective band structure from supercell calculations on alloys and impurities. *Phys. Rev. B* **85**, 085201 (2012).

**Supplemental Information**

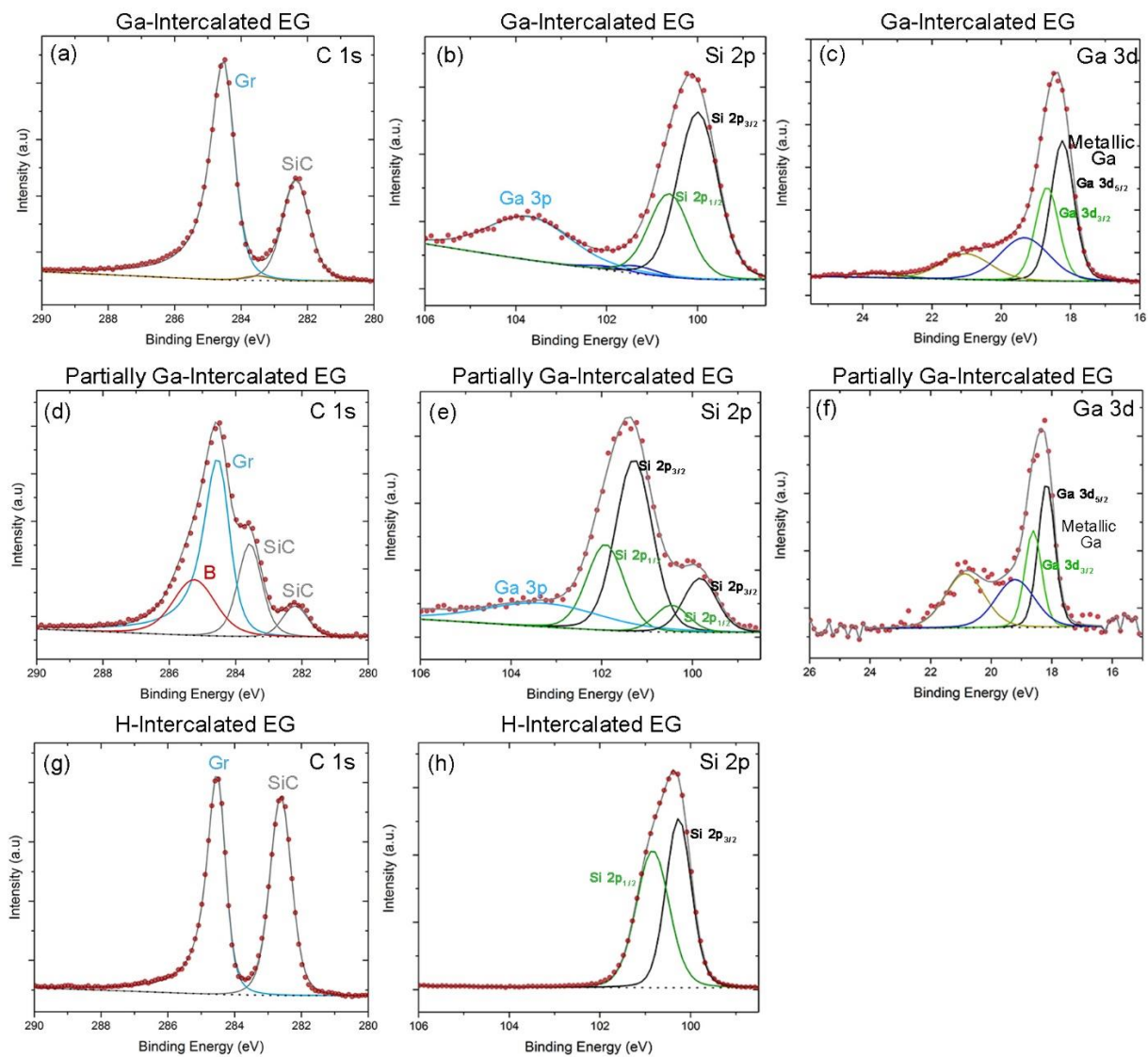

S1: X-ray photoelectron spectroscopy spectra for Ga-intercalated samples where EG is plasma-treated prior to intercalation (a-c) and where EG is not plasma-treated prior to intercalation(d-f). (g,h) Show H-intercalated samples where EG is not plasma-treated prior to intercalation. H-intercalated samples serve as a standard reference. The C 1s line shape used to fit the EG Gr peak is derived from the C 1s spectrum in (g) and S2(d).

Figure S1 shows C 1s, Si 2p, and Ga 3d core levels for different intercalated samples. Panels (a-c) show spectra acquired from a standard EG/Ga/SiC sample, where EG is exposed to an $O_2$/He plasma prior to intercalation. Ga is then intercalated at standard conditions of 800°C and 300 Torr. Intercalation leads to a shift in the C 1s peak for SiC and the Si 2p peak by 1.4-1.5 eV. A small peak near 283.5 eV may be fitted to the C 1s spectrum in (a), and is believed to correspond to a small portion of C in graphene layers that remains bonded to SiC. Ga-intercalated samples also show metallic Ga 3d peaks in the Ga 3d region which could correspond to $Ga_2O_3$ (at 21 eV), and $GaO_x$ or Ga-Si (at 19.3 eV). The spectra shown in (d-f) correspond to a Ga-intercalated sample in which EG was not exposed to an $O_2$/He plasma prior to Ga intercalation. As a result, Ga intercalation does not occur uniformly across all EG/SiC terraces. Thus, the sample is referred to as partially-intercalated, and contains island-like regions of intercalated Ga. As a result of inhomogeneous Ga intercalation, the 200μm acquisition area reflects a mixture of EG still containing a buffer layer that is bonded to SiC, as well as EG that is decoupled from SiC via intercalated Ga. The C 1s SiC peak in (d) at 283.6 eV is hypothesized to correspond to the former case, and the C 1s SiC peak at 282.2 eV to the latter. This heterogeneous surface is also reflected in the Si 2p region in (e), where two sets of Si 2p peaks are observed (one at 101.3 eV and 101.9 eV, and one at 99.8 eV and 100.5 eV). The Ga 3d region collected from this sample shows peaks similar to those in (c), however, the higher binding energy peaks at 19.2 eV and 20.9 eV are more intense relative to metallic Ga 3d peaks than those in (c). C 1s and Si 2p spectra are also shown for a reference H-intercalated EG sample (in which EG is not exposed to an $O_2$/He plasma prior to intercalation). H intercalation also results in a shift in the SiC C 1s and Si 2p peaks by ~1 eV. Because spectra are charge referenced to the $sp^2$ C (graphene) peak at 284.5 eV, relative changes in the graphene peak position are not investigated.

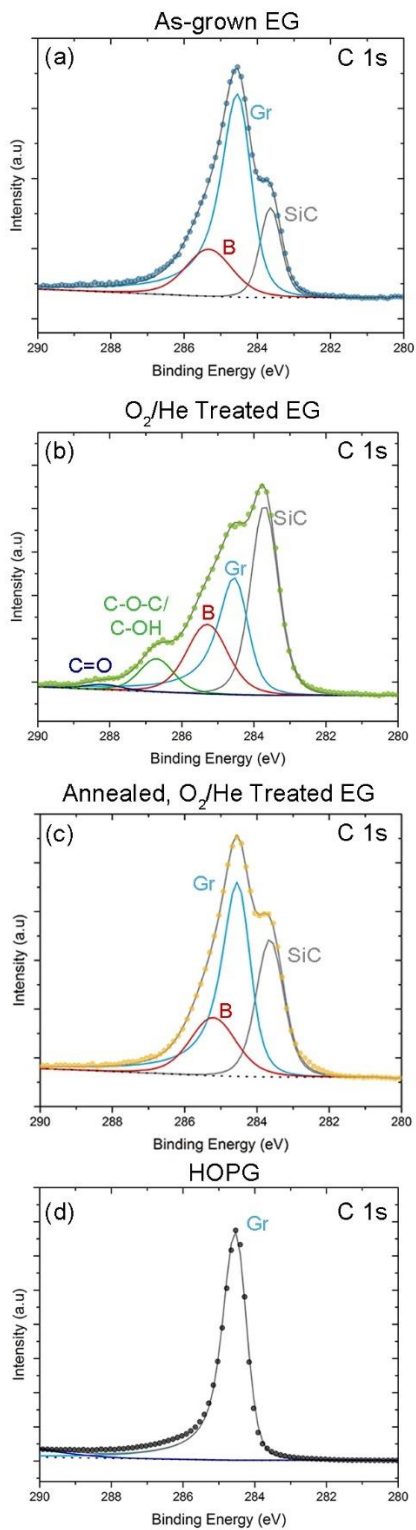

Figure S2 shows reference C 1s spectra for as-grown EG (a) as well as $O_2$/He plasma-treated EG (b). Spectra in (c) correspond to an $O_2$/He plasma-treated sample which was subsequently annealed at 800°C and 300 Torr for 30 minutes in Ar. These conditions are identical to the standard conditions used to prepare Ga-intercalated samples. This anneal was performed to investigate the stability of C–O–C/C–OH, and C=O species through the final intercalation step of the CHet process. Following this anneal, the higher binding energy peaks observed in (b) (at 286.7 and 288.2 eV) are no longer observed. This indicates that C-O-C/C-OH and C=O species do not persist throughout the intercalation step. (d) shows the C 1s region of exfoliated, highly-ordered pyrolytic graphite. This spectrum as well as that of H-intercalated EG (S1(g)) were used to derive the graphene line shape used to fit the graphene peak of the XPS spectra in these studies.

S2: XPS spectra corresponding to: As-grown EG (a), plasma-treated EG (b), plasma-treated EG which was subsequently annealed in Ar at 800°C for 30 minutes (c), and a highly-ordered pyrolytic graphite reference (d).

## Theoretical Modeling of Grapheme Defects and Single-Atom Ga-metal Intercalation

### Plasma-assisted defect formation

DFT calculations were conducted to unveil the mechanism of plasma-induced defect formation in graphene and to elucidate the interplay between EG defects and Ga metals. Following the structures used by Fampiou et al. for Pt/graphene systems,[1] a pristine graphene sheet containing 72 carbon atoms was modeled using a $6 \times 6$ hexagonal supercell with the dimensions of $14.76 \times 14.76 \times 20$ A$^3$. Subsequently, eight representative defect models, mono- to octa-vacancies, were built by the detachment of the carbon atoms from the center of the pristine supercell. The models that we considered in this study were divided into the "bare" defects, O–, and OH– passivated defects. The plasma-etching process that generates graphene defects consists of two elementary steps: vacancy creation by the bombardment of the graphene with the plasma gas which leads to the generation of dangling bonds, and adsorption of reactive plasma-components like oxygen-atoms and OH– radicals by under-coordinated carbon atoms. In this context, the formation energy of the plasma-induced defect can be considered as a sum of carbon vacancy formation energy, $E_{form,vac}$, and adsorption energy of ligands, $E_{ads}$, as also described by Geonyeop, et al.[2] where $E_{pristine}$ is the total energy of pristine layer. $E_{bare}$, $E_{O-func}$, $E_{OH-func}$ $E_{O2}$, and $E_{OH}$ are the total energies of graphene with bare, O– and OH– passivated defects, and $O_2$ and OH in a vacuum, respectively. n is number carbon atoms detached from pristine network and $\mu$ is the chemical potential of a carbon atom.

(1) $E_{form,vac} = (E_{bare} + n.\mu - E_{pristine})$

(2) $E_{ads} = (E_{O-func} - (E_{bare} + 1/2E_{O2})$ or $(E_{OH-func} - (E_{bare} + E_{OH})$

(3) $E_{form,def} = E_{form,vac} + E_{ads}$

Figure S3(a-v) shows the optimized models with bare and plasma-treated defects, and their formation energies. After the structural relaxation of the bare defects, a bond reconstruction was observed between the low-coordinated C atoms, resulting in a five-membered ring formation as a consequence of Jahn-Teller distortion that is an effect stabilizing the defects by lowering symmetry and energy.[3,4] In di- and tetra-vacancy defect models, all the dangling bonds were passivated by means of a C–C bond reconstruction that yielded 5-8-5 (Figure S3(n)) and 3-fold (Figure S3(o)) symmetric patterns, respectively. In case of saturating the under-coordinated C atoms with O– and OH–ligands, in-plane C–O–C (Figure S3(v)), and out-of-plane C=O (Figure S3(h)) and C–O–H (Figure S3(m)) bonds were formed in the models. The existence of C–O–H bond stabilizes the defects by decreasing the defect formation energy while C=O bond resulted in a less stable defect.

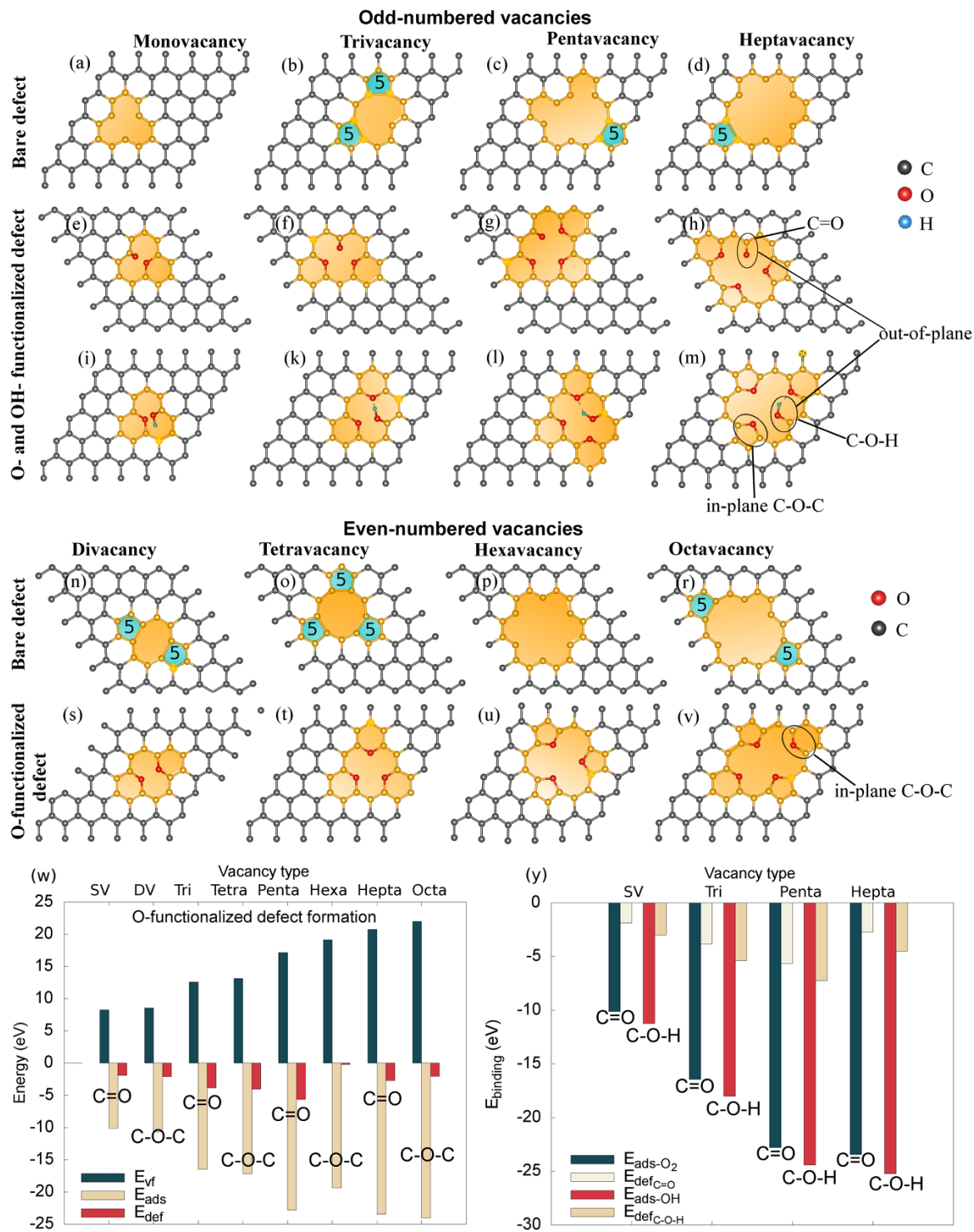

S3: (a-v) Ball-and-stick representation of the graphene networks with bare (a-d, n-r), O= and O− (e-h, s-v), and OH− (i-m) passivated defects. (w) O-functionalized defect formation ($E_{def}$), adsorption ($E_{ads}$) and carbon-vacancy (bare defect) formation ($E_{vf}$) energies depending on the incorporated functional group C–O–C and C=O. (y) O-passivated odd-numbered defect ($E_{defC=O}$) and OH-passivated odd-numbered defect ($E_{defC-O-H}$), adsorption ($E_{ads-O2}$, $E_{ads-OH}$) energies depending on the incorporated functional group C=O and C–OH. Defect types are monovacancy (SV), divacancy (DV), trivacancy (Tri), tetravacancy (Tetra), pentavacancy (Penta), hexavacancy (Hexa), heptavacancy (Hepta) and octavacancy (Octa). Odd-numbered defects are SV, tri, penta and heptavacancy.

**The Role of Plasma-induced defects on binding Ga atoms to graphene surface**

Subsequent to the investigation of the plasma-treated defect formation in graphene, the binding energies of Ga atom to the defect sites were examined. A Ga atom was supported on each optimized graphene sheets with the bare and O– and OH– passivated defects, separately. This resulted in twenty models that were allowed further structural relaxation. Figure S4(a-v) illustrates the optimized structures. The adsorption energy, $E_{ads}$, of Ga atom, was computed based on Eq. 4 where $E_{graphene+Ga}$ is the total energy of the graphene with a Ga atom adsorbed. $E_{graphene}$ and $E_{Ga}$ are the energies of the graphene sheet and an isolated Ga atom in a vacuum, respectively.

$$E_{ads} = E_{Ga+graphene} - (E_{graphene} + E_{Ga}) \tag{4}$$

As depicted in Figure S4(w, y), the adsorption of Ga atom by graphene sheet is an exothermic process, and the existence of the bare defect enhances the binding strength of Ga atom to the graphene layer as a consequence of the increase in the number of the under-coordinated edge-carbon atoms surrounding the defects (Figure S4(a-d, j-m, w)). These results reveal that the bare defect can strongly bias trapping Ga atoms in graphene by forming a covalent bond with Ga atom with a quite high binding energy (-7.38 eV). However, this, in turn, may cause an inability to release the desorbed Ga atoms from these defects since breaking the C–Ga bond would require a high dissociation energy. On the other hand, passivating the dangling bonds around the edge of the bare defects with O or OH groups significantly weakens the binding strength of Ga atom to the graphene. O-passivated defects still show higher Ga binding energies than that of pristine graphene (>1.75 eV) as illustrated in Figure S4. This indicates that defects still serve as binding sites for Ga atoms and may also allow de-trapping of Ga atoms from the graphene sheet with relatively low dissociation energies. The behavior of Ga binding to the plasma-treated graphene network also shows a discrepancy in terms of the O-induced bond and vacancy type: the odd-numbered vacancy defects (Figure S4(n-r)) bind Ga more strongly than the even-numbered ones (Figure S4(e-i)) owing to the existence of out-of-plane C=O double bonds which act as trapping centers for the Ga atom (Figure S3(h) and Figure S3(w-I, y-I)). Contrary to the carbonyl (C=O) bond formation, the ether (C–O–C) groups (where carbon atoms exhibit an $sp^2$ hybridized form) are capable of contributing to the stabilization of the even-numbered defects by the pair-wise removal of the unsaturated bonds (Figure S3(v) and Figure S4(w-II). There is also an evident trend between the vacancy size and the binding strength of the Ga atom. As depicted in Figure S4(w-II), octa- and hepta-vacancy defects have highest binding energies of -2.08 and -3.36 eV, respectively, among the even- and odd-numbered vacancy types, indicating that the vacancy size plays a crucial role in tuning the defect/metal interaction, and can enable control over Ga intercalation. When the edge atoms in question are saturated with OH– groups in the odd-numbered defects (Figure S4(s-v)), the C–O–H bond formation dramatically alters the binding strength of Ga atom, and can result in lower binding energy even than that of the pristine network (<1.75 eV). As depicted in Figure S4(t, u) and Figure S4(y-II), tri and penta-vacancy defects saturated with OH– groups only weakly attract the Ga metal atoms, with the energies of -1.41 and -1.35 eV, respectively. This specifies that the defects saturated with OH– are unable to draw Ga atoms to the surface and may cause a clustering between the Ga metals, and such a weak interaction between graphene and metal atoms may have a detrimental effect on Ga intercalation. Note that mono and hepta-vacancies in Figure S4(s,v) have higher binding energies than pristine graphene since in these cases the Ga atom interacts with O atoms instead of the H atom.

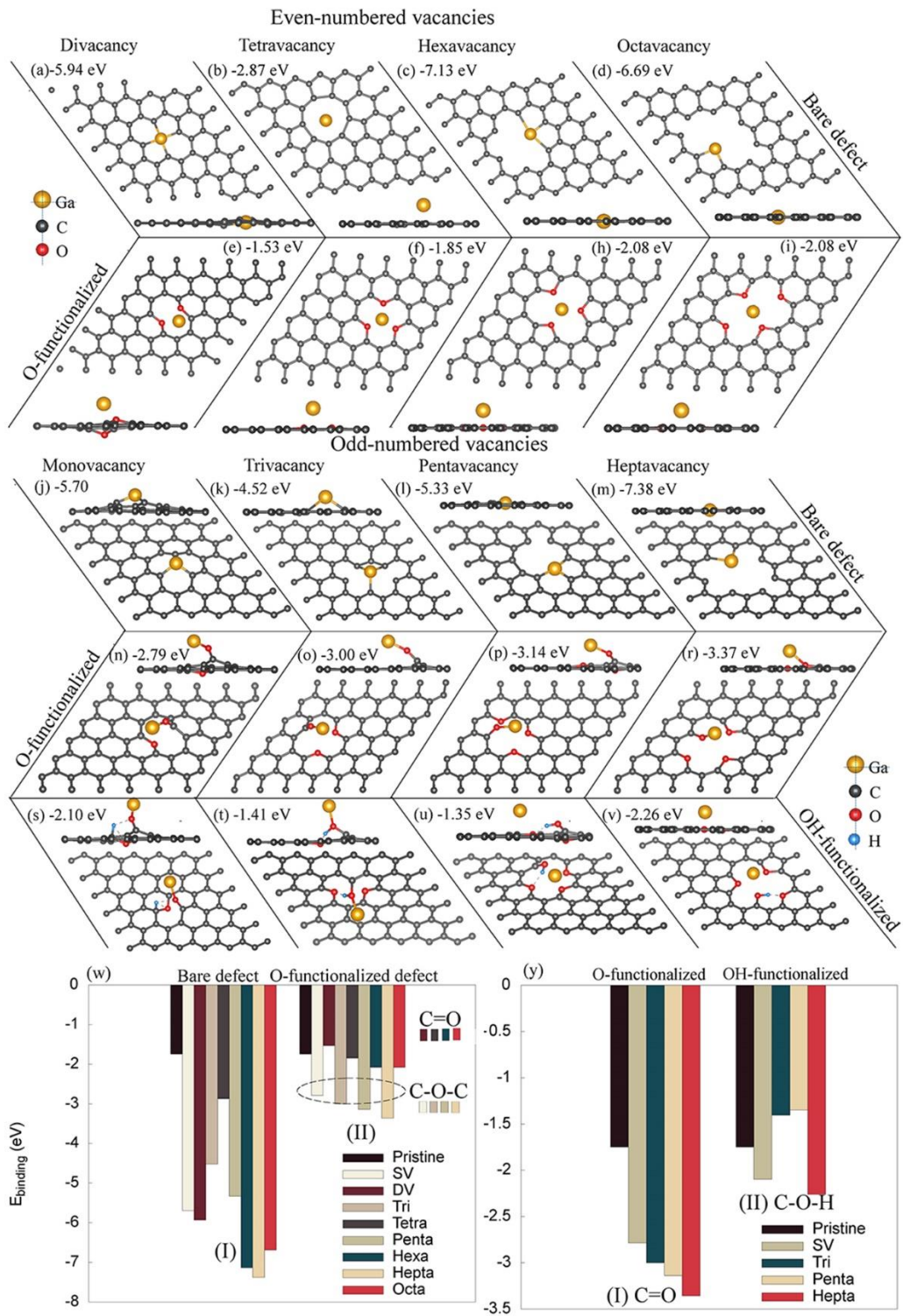

S4: Ball-and-stick representation of Ga atom adsorbed on graphene networks with (a-d, j-m) bare defects, (c-i, n-r) O- and (s-v) OH−passivated defects where their individual binding energies are presented. (e-i) correspond to C=O passivation and (n-r) correspond to C−O−C passivation. (w, y) Binding energies of Ga metal to defective graphene with/without the functional groups where (w-I) corresponds to bare defects, (w-II) bars with light and dark colors correspond to C−O−C and C=O, respectively, (y-I) corresponds to C=O, and (y-II) corresponds to C−O−H. Defect types are monovacancy (SV), divacancy (DV), trivacancy (Tri), tetravacancy (Tetra), pentavacancy (Penta), hexavacancy (Hexa), heptavacancy (Hepta) and octavacancy (Octa).

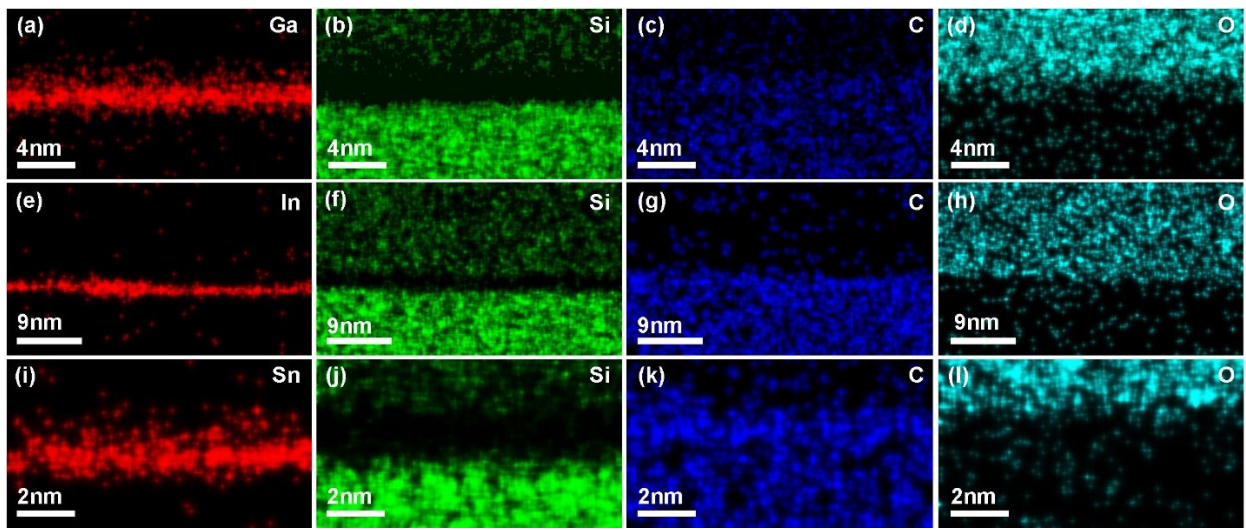

S5: Energy dispersive spectroscopy maps collected for Ga, In, and Sn intercalated samples. Oxygen signal is located above intercalant layers, indicating the metal films are not oxidized.

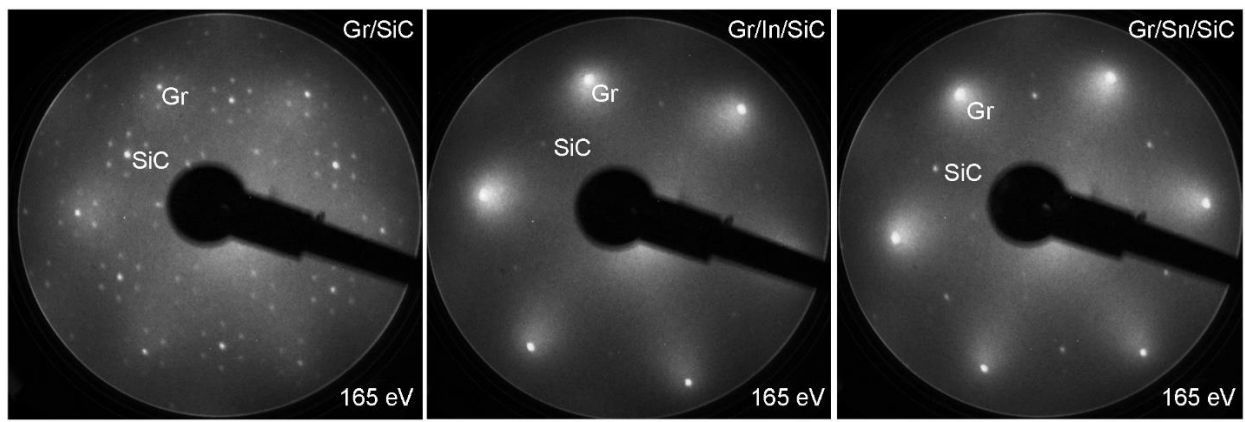

S6: Low energy electron diffraction patterns for Gr/SiC, Gr/In/SiC, and Gr/Sn/SiC acquired at 165eV. Additional spots in Gr/SiC pattern correspond to buffer layer reconstruction

The phase stabilities of 1L-, 2L-, and 3L-metals (in red, green, and blue lines) discussed in the main text and in Figure 2b do not include a bilayer graphene cap. The cases when bilayer graphene is included are shown in Fig. S7. For Ga and In the results are qualitatively unchanged: the allowed range of metal chemical potentials would yield 1, 2, or 3 layers of Ga and 1 or 2 layers of In. For Sn, although the trilayer stabilizes itself against the monolayer structure near the bulk Sn chemical potential by relaxing into a simple-hexagonal $Sn_{Si}Sn_{Si}Sn_{Si}$ stacking, the higher-energy bilayer structure within the same chemical potential range (relaxed into a distorted structure where Sn atoms the second Sn layer are no longer coplanar) may kinetically prevent the system from accessing the trilayer structure. This is consistent with the blurred STEM images of the second Sn layer discussed in the main text.

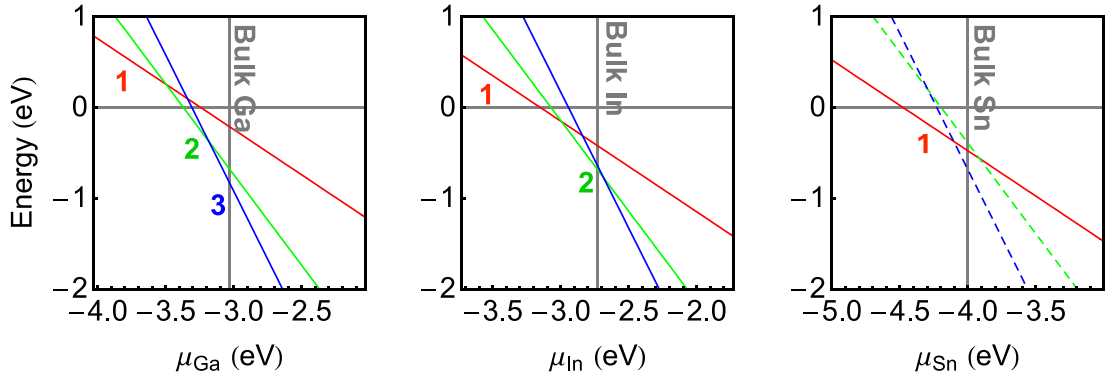

Figure S7. Phase Stability, including bilayer graphene cap. Same as in Fig. 2b, phase stability of 1L-, 2L-, and 3L-metals as functions of metal chemical potentials, but now including a bilayer graphene cap.

Figure S8 shows differing metal layer numbers between EG and SiC. The layer numbers displayed in (a), (c), and (d) are observed only occasionally, in select regions of sample cross-sections. These different thicknesses are likely due to nearby step edges, such as those shown in Figure S9. (b) shows a cross-sectional STEM image of 2L-Ga, which is frequently observed along with 3L-Ga. (e) and (f) demonstrate the mixture of 1 and 2 metal layers observed following Sn intercalation.

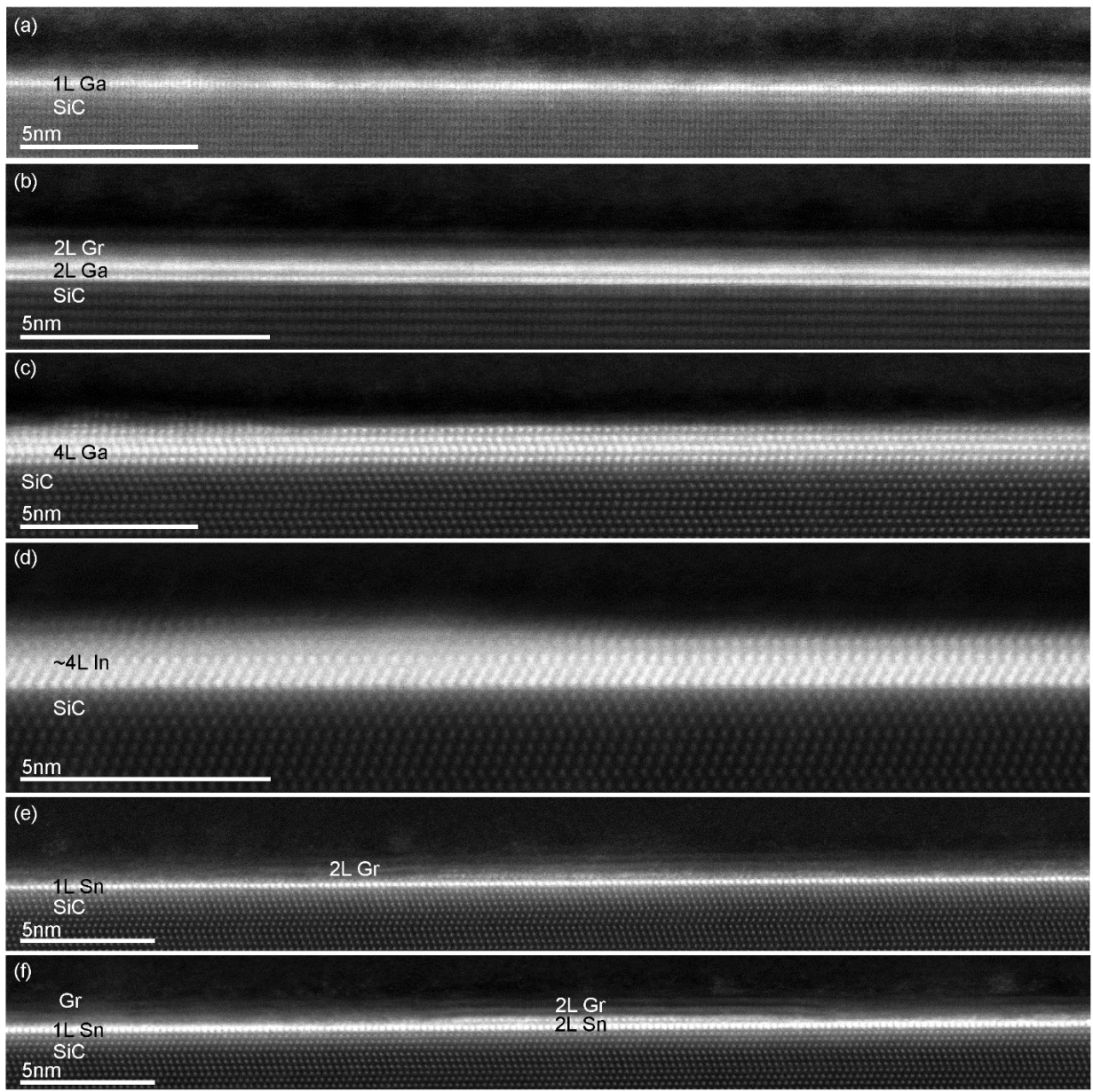

S8: Cross-sectional STEM images showing additional observed thicknesses of intercalated Ga, In, and Sn layers, ranging from 1-4. The predominantly observed layer numbers are 2-3 Ga layers, 2 In layers, and 1 Sn layer. Graphene layer number above metals is 1 for (a) and (c), and 2 for (b-f).

To understand the impact of SiC step edges on intercalant layers, samples are investigated with cross-sectional STEM. Cross-sectional STEM images in Figure S9 show intercalated indium layers over a length of approximately 600 nm. Within a SiC terrace region (c), uniform bilayer In is observed, however, step edges in the SiC clearly impact the In film, where step edges of approximately one atomic layer high can result in an increased layer number of intercalated species (d), and step edges greater than a few atomic-layers in height can disrupt the film (b). Regions of continuous, bilayer In films have also been observed across single atomic-level steps.

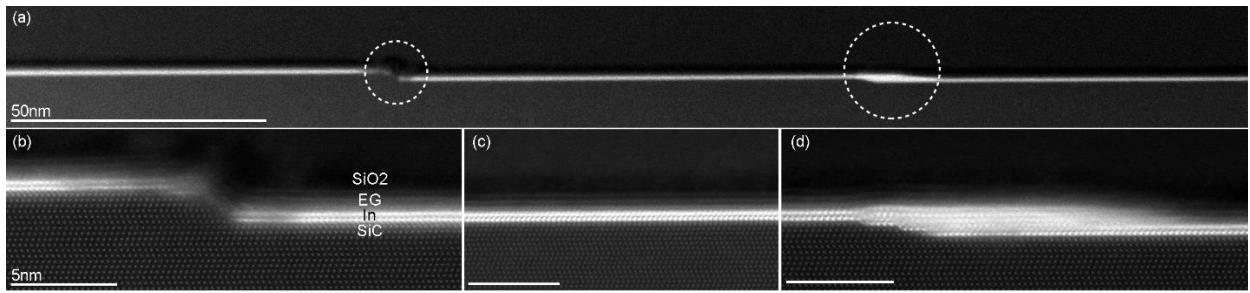

S9: (a) Cross-sectional STEM of Gr/In/SiC structures, where step height is directly shown to impact nearby In layer number. Circles in (a) correspond to regions (b) and (d). (c) shows the region of In between these two circles, well within a terrace region, where step edges do not impact metal layer number.

Auger maps of intercalated-Ga and In samples are consistent with cross-sectional STEM observations (Figure S9), where strong C and Ga/In signal is observed across the 10x10μm square area. The diagonal lines in (a-c, e-g) correspond to step edges in the SiC which can yield additional graphene layers. The increased C signal at the large diagonal features in (b) and (f) is consistent with a greater number of EG layers, and is accompanied by reduced Ga and In signal (a,e), which is likely due to signal attenuation from a greater number of overlying graphene layers. These step edges may contain several additional layers of graphene, compared to other step edges, such as those indicated by dashed lines in (a), which may be smaller in height, similar to those in Figure S9(d).

Regions with increased oxygen signal in (d) and (h) can be attributed in part to small metal-oxide islands that have nucleated on top of graphene layers near step edges. These regions show both high oxygen and metal signal, as shown in the small circular features in (a) and (d). However, some high-oxygen regions correspond to regions with decreased metal signal (left-hand region in (a) and (d), bottom center region in (e) and (h)). These regions could contain a silicon oxide and/or fewer graphene layers. The chemistry of these regions is still under investigation.

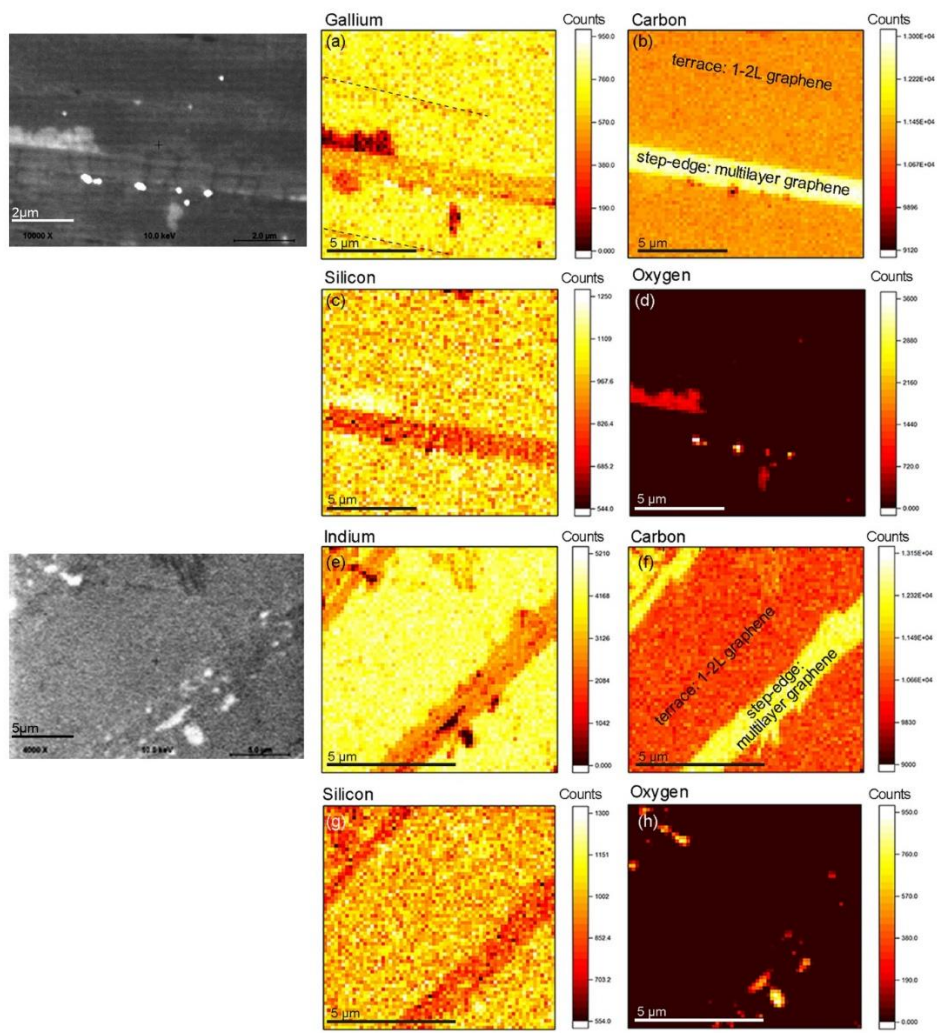

S10: Auger electron spectroscopy maps of Gr/Ga/SiC and Gr/In/SiC samples and corresponding scanning electron microscope images. Ga, In, C, Si, and O are shown in the above maps, which display terrace and step-edge regions for both samples. Step-edges display stronger relative C signal, which could indicate that greater numbers of graphene layers attenuate the signal of the underlying Si and metal layers. Small O-rich regions are observed near step edges, where some metallic islands have nucleated on top of the graphene layers.

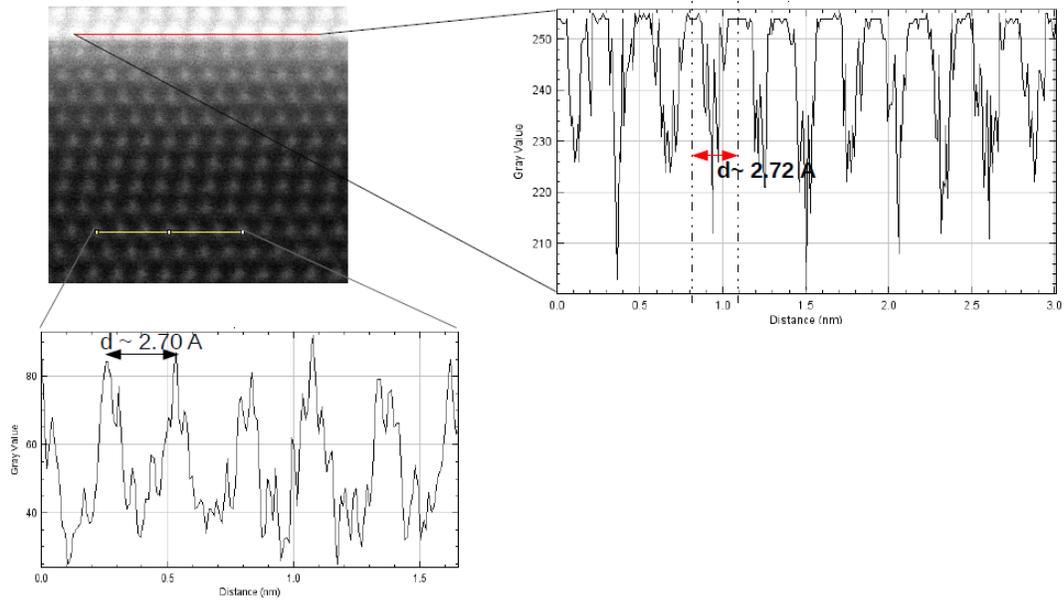

S11: Cross-sectional STEM image of Ga/SiC, where SiC atoms across the yellow line show a spacing of 2.70Å, and the Ga atoms across the red line show a spacing of 2.72Å

**Ga Stacking order at the bilayer graphene/SiC interface**

The thermodynamic ground states for bilayer and trilayer Ga are discussed in the main text without including a capping bilayer graphene. The relative energies of all possible bilayer structures are listed in Table S1 under "w/o graphene cap", where Site 2 lies further away from the SiC surface than Site 1. By including a bilayer graphene cap (necessitating a larger 2×2 bilayer graphene + √3×√3 R30º Ga/SiC supercell), the order of the relative energies is not altered, as shown under "with graphene cap". The same applies for the trilayer case, with the following exception. The $Ga_CGa_{hollow}Ga_{hollow}$ and $Ga_{hollow}Ga_{Si}Ga_{Si}$ structures become unstable when the capping bilayer graphene is added and transform to $Ga_{Si}Ga_{hollow}Ga_C$ and $Ga_{Si}Ga_CGa_{Si}$ respectively. The energies of the original unstable structures (marked by asterisks) are estimated using a force convergence threshold (0.05 eV/Å) larger than that enforced for every other case (0.01 eV/Å). Even with these exceptions the ground state is still $Ga_{Si}Ga_CGa_{hollow}$.

| Site 1 | Site 2 | w/o graphene cap | with graphene cap |
|--------|--------|------------------|-------------------|
| C | Si | 0.56 | 0.54 |
| C | C | 0.51 | 0.49 |
| C | Hollow | 0.47 | 0.45 |
| Hollow | Si | 0.45 | 0.43 |
| Hollow | Hollow | 0.42 | 0.42 |
| Hollow | C | 0.3 | 0.30 |
| Si | Si | 0.22 | 0.23 |
| Si | Hollow | 0.05 | 0.05 |
| Si | C | 0 | 0 |

| Site 1 | Site 2 | Site 3 | w/o graphene cap | with graphene cap | local instability |
|--------|--------|--------|------------------|-------------------|-------------------|
| C | Hollow | Hollow | 0.58 | 0.58* | 0.02 (→Si hollow C) |
| Hollow | Si | Si | 0.49 | 0.50* | 0.04 (→Si C Si) |
| C | Hollow | Si | 0.46 | 0.45 | |
| Hollow | Si | C | 0.35 | 0.36 | |
| Si | C | C | 0.04 | 0.04 | |
| Si | C | Si | 0.03 | 0.02 | |
| Si | C | Hollow | 0 | 0 | |

Table S1. Relative stability of possible stacking orders of bilayer and trilayer Ga, following the notation of $Ga_{Site1}Ga_{Site2}Ga_{Site3}\dots$ with site indices increasing further away from the Si/Ga interface. See text for the discussion on locally unstable structures with energies marked by asterisks.

**Details on band structures at the PBE level and hybrid functional level**

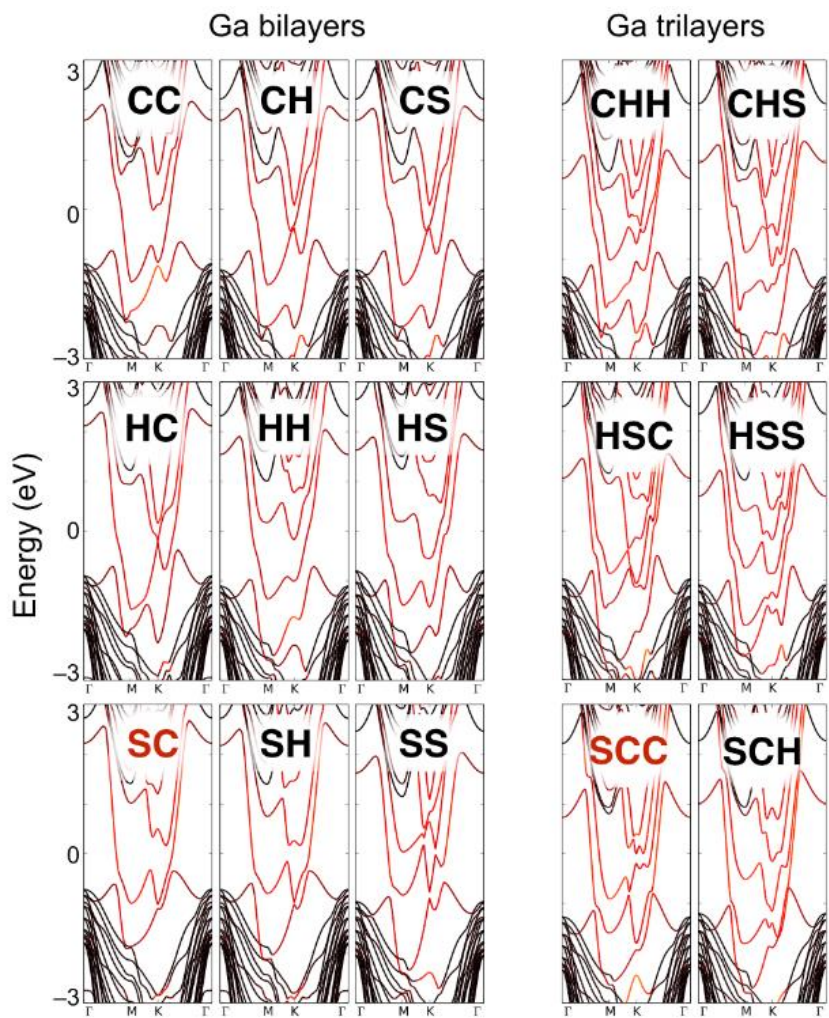

S12: Projected band structures for all possible bilayer Ga geometries and for selected trilayer Ga geometries calculated at the DFT level, where Ga orbital characters are indicated in red. The best matching case for bilayer and trilayer are $Ga_{Si}Ga_C$ and $Ga_{Si}Ga_CGa_C$ (highlighted in red). This is consistent with $Ga_{Si}Ga_C$ being thermodynamic ground state for the bilayer case and $Ga_{Si}Ga_CGa_C$ being the nearly-degenerate next-lowest-energy structure for the trilayer case.

Figure S12 shows calculated band structures for bilayer and trilayer Ga on SiC. These calculations are compared with experimental ARPES measurements (main text Figure 3) to find the most favorable 2D-Ga structures. Among the bilayer band structures, the $Ga_{Si}Ga_C$ case achieve the best agreement in terms of relative band positions, with the only exception that the Fermi energy appears to be off by 0.6 eV. For trilayers, the $Ga_{Si}Ga_CGa_C$ and $Ga_{Si}Ga_CGa_{hollow}$ band structures both show some deviations near K but match with additional bands with weak intensities in ARPES and have Fermi level in alignment with the ARPES measured one. Two other geometries with extra carbon atoms near the SiC/Ga interface were also considered but gave drastically different band structures.

For the best matching cases, bilayer $Ga_{Si}Ga_C$ and trilayer $Ga_{Si}Ga_CGa_C$, we performed additional band structure calculations at the hybrid functional level (HSE06, see Methods) to rule out the possibility that the above band structure deviations could be due to the intrinsic delocalization error of approximate functionals at the DFT level. As shown in Figure S13, we observe an overall energy rescaling that increases the bandwidth of the metal by expanding states away from the Fermi level. For the case of bilayer Ga, the leftmost band crossing point along Γ–M lowers away from the Fermi level, from –0.6 eV in the DFT (PBE functional) case to –1.1 eV in the hybrid functional case. The latter energy separation matches with the ARPES measured one (–1.2 eV) better, thus removing the need to impose an artificial Fermi level shift as discussed in the main text. Thus we conclude that the dominant surface phase is bilayer $Ga_{Si}Ga_C$ geometry, consistent with it being the ground state of bilayer Ga, likely with co-existing $Ga_{Si}Ga_CGa_C$ structures.

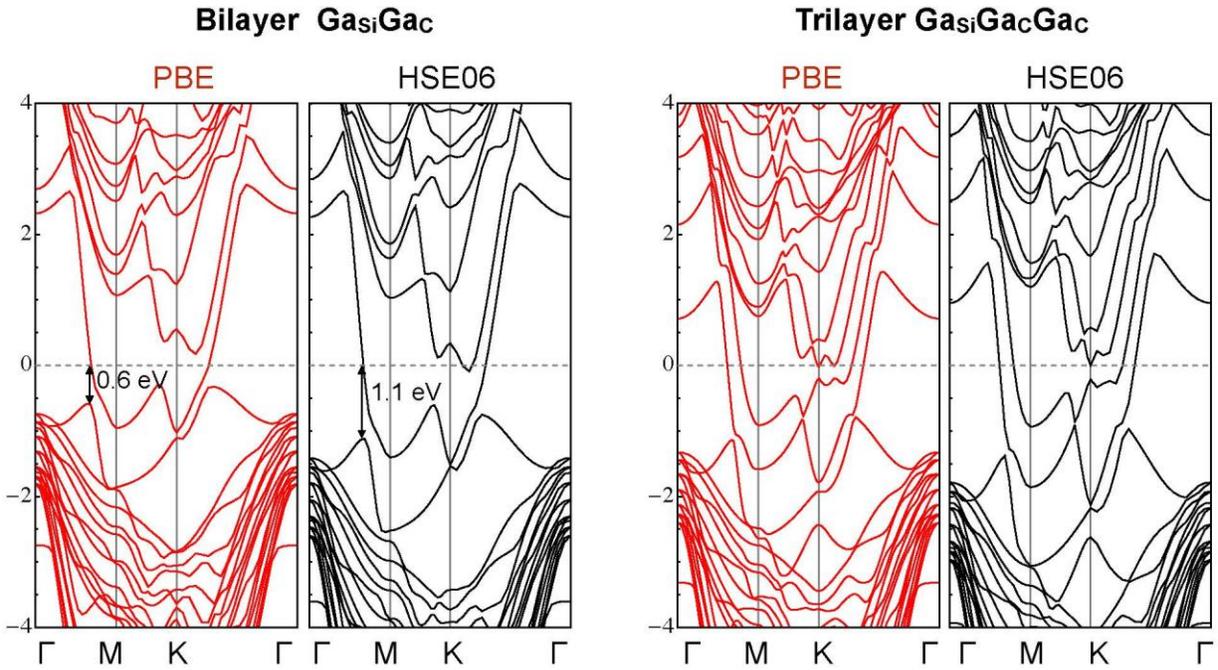

Figure S13. Band structure for the (left) $Ga_{Si}Ga_C$ bilayer and the (right) $Ga_{Si}Ga_CGa_C$ case calculated at DFT and hybrid functional (HSE06) levels.

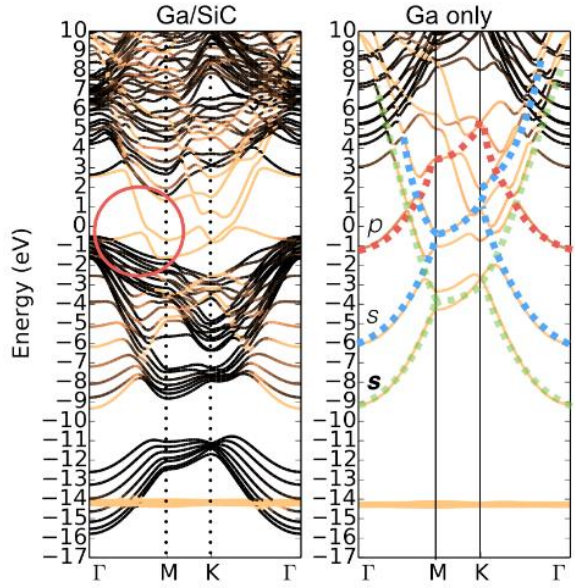

S14: DFT band structure of bilayer Ga/SiC without graphene. Comparing with the bands of hypothetical freestanding bilayer Ga (right panel), the origin of the bands with Ga orbital characters colored in orange can be assigned $s$ bonding, $s$ antibonding, and $p$ characters. The $s$ band with the deepest level origin contributes most to the Fermi surface.

The approach used for ARPES discussion accompanying Figure 3 in the main text is as follows: we construct a 2×2 graphene + √3×√3 R30º Ga/SiC supercell; its deviation from the ideal 13×13 graphene + 6√3×6√3 R30º Ga/SiC supercell induces an 8% artificial strain to the graphene lattice and a consequential ~0.5 eV increase in its work function.[5] Thus we only compare selected band features with ARPES in the first approach, whereas band alignment between Ga and graphene could be off by 0.5 eV. For the second approach, we construct a 5×5 graphene + 4×4 R0º Ga/SiC supercell. Although the relative interfacial orientation is incorrect, this supercell avoids the creation of the artificial interfacial strain and should yield more accurate charge transfer and band alignments. The resulting doping level of graphene for bilayer and trilayer Ga are 0.15 and 0.42 eV, consistent with the work function variation between the two: 4.61 and 4.06 eV for bilayer and trilayer Ga. Thus it appears that whereas band features more closely resemble the calculated bands for bilayer Ga, the band alignment and filling suggests the presence of trilayer Ga.

To reveal the orbital origin of the band crossing along ΓM$_{Ga}$ in Figure 3a,b, we compare the projected band structure of bilayer Ga/SiC (without graphene) to a hypothetical freestanding bilayer Ga where Ga atoms are frozen at their positions in the hybrid system, as shown in Figure S14. The latter clearly shows three nearly-free-electron-like bands of $s$-bonding, $s$-antibonding, and $p$ orbital character. The band crossing is thus hybridization between a parabolic $s$ orbital originating from ~9 eV below the Fermi level and the $p$ orbital near the Fermi level. Similarly, the band crossings along ΓK$_{Ga}$ are also between $s$ and $p$.

To verify whether the 2D Ga is under in-plane epitaxial strain, we calculated the Ga contribution to the total strain energy of the hybrid system by subtracting the contribution of the bare SiC substrate from the total. The minimum strain energy occurs at 95% and 96% of the in-plane lattice constant of SiC (0001) for bilayer and trilayer Ga. Thus the Ga region is under moderate tensile strain.

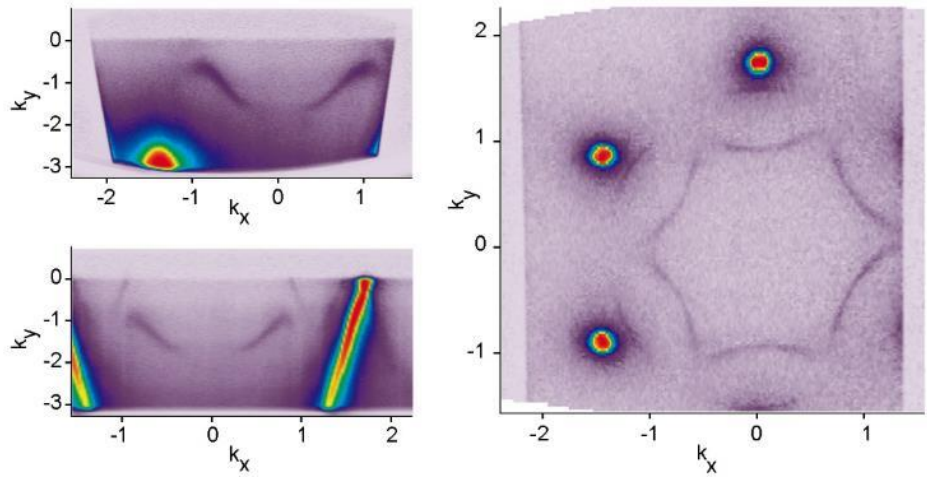

S15: ARPES measurements of Gr/In/SiC, showing similarities to Gr/Ga/SiC measurements in the form of graphene bands, avoided crossing points, and an In/SiC BZ rotated 30° from graphene.

Additional references for methods section are references 6-20 in the bibliography shown below.

## References

1. Fampiou, I. & Ramasubramaniam, A. Binding of Pt Nanoclusters to Point Defects in Graphene: Adsorption, Morphology, and Electronic Structure. *J. Phys. Chem. C* **116**, 6543–6555 (2012).

2. Lee, G., Kim, J., Kim, K. & Han, J. W. Precise control of defects in graphene using oxygen plasma. *J. Vac. Sci. Technol. A Vacuum, Surfaces, Film.* **33**, 060602 (2015).

3. Banhart, F., Kotakoski, J. & Krasheninnikov, A. V. Structural Defects in Graphene. *ACS Nano* **5**, 26–41 (2011).

4. Ronchi, C. *et al.* π Magnetism of Carbon Monovacancy in Graphene by Hybrid Density Functional Calculations. *J. Phys. Chem. C* **121**, 8653–8661 (2017).

5. Choi, S.-M., Jhi, S.-H. & Son, Y.-W. Effects of strain on electronic properties of graphene. *Phys. Rev. B* **81**, 081407 (2010).

6. Subramanian, S. *et al.* Properties of synthetic epitaxial graphene/molybdenum disulfide lateral heterostructures. *Carbon N. Y.* **125**, 551–556 (2017).

7. Bersch, B. *et al.* An Air-Stable and Atomically Thin Graphene/Gallium Superconducting Heterostructure. *https://arxiv.org/abs/1905.09938* (2019).

8. Giannozzi, P. *et al.* QUANTUM ESPRESSO: a modular and open-source software project for quantum simulations of materials. *J. Phys. Condens. Matter* **21**, 395502 (2009).

9. Joubert, D. From ultrasoft pseudopotentials to the projector augmented-wave method. *Phys. Rev. B - Condens. Matter Mater. Phys.* **59**, 1758–1775 (1999).

10. Blöchl, P. E. Projector augmented-wave method. *Phys. Rev. B* **50**, 17953–17979 (1994).

11. Perdew, J. P., Burke, K. & Ernzerhof, M. Generalized Gradient Approximation Made Simple. *Phys. Rev. Lett.* **77**, 3865–3868 (1996).

12. Perdew, J. P., Burke, K. & Ernzerhof, M. Erratum: Generalized gradient approximation made simple (Physical Review Letters (1996) 77 (3865)). *Phys. Rev. Lett.* **78**, 1396 (1997).

13. Marzari, N., Vanderbilt, D., De Vita, A. & Payne, M. C. Thermal Contraction and Disordering of the Al(110) Surface. *Phys. Rev. Lett.* **82**, 3296–3299 (1999).

14. Stukowski, A. Visualization and analysis of atomistic simulation data with OVITO–the Open Visualization Tool. *Model. Simul. Mater. Sci. Eng.* **18**, 015012 (2010).

15. Momma, K., Izumi, F. & IUCr. *VESTA 3* for three-dimensional visualization of crystal, volumetric and morphology data. *J. Appl. Crystallogr.* **44**, 1272–1276 (2011).

16. Makov, G. & Payne, M. C. Periodic boundary conditions in ab initio calculations. *Phys. Rev. B* **51**, 4014–4022 (1995).

17. Neugebauer, J. & Scheffler, M. Adsorbate-substrate and adsorbate-adsorbate interactions of Na and K adlayers on Al(111). *Phys. Rev. B* **46**, 16067–16080 (1992).

18. Enkovaara, J. *et al.* Electronic structure calculations with GPAW: a real-space implementation of the projector augmented-wave method. *J. Phys. Condens. Matter* **22**, 253202 (2010).

19. Kresse, G. & Furthmu, J. Efficient iterative schemes for ab initio total-energy calculations using a plane-wave basis set ¨. *Phys. Rev. B* **54**, 11169–11186 (1996).

20. Heyd, J., Scuseria, G. E. & Ernzerhof, M. Hybrid functionals based on a screened Coulomb

potential. *J. Chem. Phys.* **118**, 8207–8215 (2003).